# Electron Counts, Structural Stability, and Magnetism in $BaCuSn_2$-$CeNi_{1-x}Si_2$-type $YT_xGe_2$ (T= Cr, Mn, Fe, Co, and Ni)

*Léa Gustin,*[a] *Lingyi Xing,*[b] *Max T. Pan,*[a] *Rongying Jin,*[b] *Weiwei Xie*[a]*

[a] Department of Chemistry, Louisiana State University, Baton Rouge, LA, 70803

[b] Department of Physics and Astronomy, Louisiana State University, Baton Rouge, LA, 70803

## ABSTRACT

Results of crystallographic refinement, the relationship between electron counts and structural stability, and magnetic characterization of $YT_xGe_2$ (T= Cr, Mn, Fe, Co, and Ni) prepared using the arc melting method are presented. These $YT_xGe_2$ compounds crystallize in the $BaCuSn_2$-$CeNi_{1-x}Si_2$-type structure with space group *Cmcm*, and the site occupancies of 3*d* transition metals range from $x$ = 0.22(1) for Cr to $x$ = 0.66(1) for Ni. Based on a combination of single crystal and powder X-ray diffraction and scanning electron microscopy, the trends are clearly established that the smaller transition metal atoms exhibit larger occupancies on T (Cu) site. Our investigation into the relationship between electron count and site defect reveals that a stable configuration is obtained when reaching 10.3e- per transition metal (Y+T), which strongly correlates with the defect observed in the case of T metals. Magnetic properties measurements indicate paramagnetism for T = Cr, Fe, and Co, but ferromagnetism for T = Mn with a Curie temperature of ~ 293 K and effective moment ~ 3.6 $\mu_B$/Mn. The absence of superconductivity in this series is surprising because they consist of similar building blocks and electron counts to superconducting $YGe_{1.5+\delta}$ except for 3*d* transition metals. Introducing 3*d* transition metals into the system plays a critical role in suppressing superconductivity, offering new insights into the interplay between superconductivity and magnetism in layered intermetallics.

## 1. Introduction

Understanding the relationship between crystal structures, chemical compositions, and physical properties of complex intermetallic compounds remains challenging because few general prediction rules allow targeted synthesis of solid-state materials with specific properties, especially superconductivity and ferromagnetism.[1,2] Given the extrapolation difficulty with balancing the physical properties and the chemical stability of the compounds, many strategies have been proposed for finding new functional materials.[3] A standard chemical view into increasing the likelihood of finding a target superconductor or ferromagnetic material is assumed to be operating in the structural family with unstable electronic structures.[4–6] One of most paramount rules for estimating the ferromagnetic properties in a known simple structure type is the Stoner criterion.[7,8] Evaluation of the Stoner condition can be expressed as $N(E_F)*I > 1$, where $N(E_F)$ is the density of states (DOS) at the Fermi level, and $I$ is the Stoner constant for a specific element. Thus, large density of states at Fermi level may increase the chance toward ferromagnetism by breaking the spin degeneracy in materials. Similarly, based on the Bardeen-Cooper-Schrieffer (BCS) theory, the qualitative interrelationship between the superconducting critical temperature ($T_c$) and $N(E_F)$ can be expressed using the equation of $k_BT_c = 1.13\ \hbar\ \omega\ \exp(-1/N(E_F)V)$, where V is related to the electron-phonon interaction and ω is a characteristic phonon frequency.[9,10] According to the expression, a large $N(E_F)$ may result in higher $T_c$. Previous studies on substituting magnetically-active 3*d* elements with non-magnetic 4*d* elements in $Hf_5Sb_{3-x}T_x$ and $Zr_5Sb_{3-x}T_x$ (T = transition metals) offers a new platform to design and synthesize new superconductors from the reported ferromagnetic materials.[11,12] Vice versa, one may explore new ferromagnetic materials from reported superconductors to understand how to tune the magnetism and superconductivity in the structural family by substituting elements.

Orthorhombic $YGe_{1.5+\delta}$ with space group *Cmmm* is a previously reported superconductor with $T_c$ ~ 3.8 K.[13] In $YGe_{1.5+\delta}$, the structure can be considered as a combination of ordered $YGe_{1.5}$ and deficient Ge layers. Adding 3*d* elements such as Fe, Co, or Ni to $YGe_2$ can produce new transition metal deficient compounds.[14,15] The products crystallize in the $BaCuSn_2$-type structure stabilized by transition metals as interstitial. However, such subtle change suppresses superconductivity and induces magnetic interactions. With these magnetic and structural features in mind, we report herein a thorough structural characterization and investigation of the magnetic properties of $YT_xGe_2$ (T= Cr, Mn, Fe, Co, and Ni) phases, with an emphasis on atomic occupancies throughout the unit cell, electron counts, and magnetic properties. In particular, $YT_xGe_2$ without *f* electrons on Y atoms will simplify the magnetic properties in the compounds without considering the interactions between *f* electrons on rare earth metals and *d* electrons from transition metals.

## 2. Material and Methods:

### 2.1 Synthesis

Polycrystalline samples of $YT_xGe_2$ (T = Cr, Mn, Fe, Co, and Ni) were synthesized by arc melting the elemental metals (total ~ 300mg) under an argon atmosphere on a water-cooled stage using a tungsten electrode. Zr was used as an oxygen getter. Samples were obtained by weighing out Y (Alfa Aesar, pieces, 99.9%) and T = Cr, Fe, Mn, Co, and Ni (Alfa Aesar, pieces, 99.8-99.97%) in a 1 : *x* ratio ($0.06 < x < 1$). They were reacted with stoichiometric amounts of Ge (Alfa Aesar, 99.9999%) with a 3% excess to compe nsate for any loss during the arc melting process. Different loading samples were prepared (see Table 1) and they all contain a minor impurity phase of $YGe_{1.5+\delta}$, which can be detected by magnetic measurements. High purity $YGe_{1.5+\delta}$ ($\delta$= 0.17) was synthesized by arc-melting Y and Ge with atomic ratio Y:Ge=1:1.67. The products were turned and melted several times to ensure good sample homogeneity. All samples are stable upon exposure to air and moisture.

### 2.2 Phase Identification by Powder X-ray Diffraction (PXRD)

Finely ground samples were examined by powder X-ray diffraction for phase identification on a Rigaku MiniFlex 600 powder diffractometer employing Cu radiation ($\lambda_{K\alpha}$= 1.5406 Å). The scattered intensity was measured with Bragg angle (2θ) ranging from 5° to 90° at a rate of 1°/min using a scintillation detector with a step of 0.01° 2θ in step scan mode. Unit cell parameters were refined through Le Bail fit using JANA2006.[16,17]

### 2.3 Crystal Structure Determination by Single Crystal X-ray Diffraction (SCXRD)

Small crystals (~0.015 mm in diameter) of each $YT_xGe_2$ (T= Cr, Mn, Fe, Co, and Ni) samples were picked out from the ground arc melted samples and mounted on a Kapton loop. Intensity data were collected on a Bruker Apex II diffractometer using Mo $K_\alpha$ radiation (λ = 0.71073 Å); data were collected with 0.5° scans in ω with an exposure time of 10s per frame. The 2θ range extended from 4° to 70°. The *SMART* software was used for data acquisition.[14] Intensities were extracted and corrected for Lorentz and polarization effects using the *SAINT* program.[19] With the *SHELXTL* package, the crystal structures were solved using direct methods and refined by full-matrix least-squares on $F^2$.[20]

### 2.4 Energy-Dispersive X-ray Spectroscopy (EDX)

Chemical composition characterization was completed using a high vacuum scanning electron microscope (SEM) (JSM-6610 LV). Samples were placed on carbon tape prior to loading into the SEM chamber. Multiple points and areas were examined for each sample to get the Y: T ratio. At a voltage of 15 kV, spectra were collected for 100 seconds to get the chemical composition via TEAM EDAX software.

### 2.5 Magnetic Measurements

Investigation of the magnetic properties of the $YT_xGe_2$ (T= Cr, Mn, Fe, and Co) compounds (temperature-dependent magnetic susceptibility and hysteresis loops) was carried out using a Quantum Design magnetic property measurement system (MPMS-7T). Each sample was measured in both zero-field-cooled (ZFC) and field-cooled (FC) modes under the application of magnetic field. M(H) curves were obtained for the T = Mn sample between 2 and 390 K.

## 3. Results and Discussion

According to previous study, $BaCuSn_2$-$CeNi_{1-x}Si_2$-type $YTGe_2$ phases have been documented for T = Fe, Co, and Ni.[14] However, no refinement of the atomic positions or the occupancies was reported leaving the question of possible defects present in the structure that our study focuses on. While there was report about partial occupancy in $YMn_{0.3}Ge_2$ and $YCo_{0.55}Ge_2$, only the former compound was thoroughly characterized structurally.[15,21] Additionally, we extend the $BaCuSn_2$-$CeNi_{1-x}Si_2$-type $YT_xGe_2$ series to the Cr analog that was once reported without any structural information.[22] This project yielded $BaCuSn_2$-$CeNi_{1-x}Si_2$-type $YT_xGe_2$ with atomic occupancies increasing from Cr to Ni. Because this structure type involves the atomic distributions of transition metals, this aspect of chemical compositions may also influence the bulk physical properties. The relationship between the atomic occupancies and atomic sizes of the 3*d* elements and electron counts per metal (Y+T) were investigated.

### 3.1 Phase Analyses

In contrast with the previously reported $YTGe_2$ compounds (where T = Mn and Co) synthesized through arc melting process, here different loading compositions were attempted to obtain the phase formation boundaries of $BaCuSn_2$-$CeNi_{1-x}Si_2$-type $YT_xGe_2$ with *x* ranging from 0.0625 to 1. Only those reactant mixtures that were loaded with $YCr_{0.2}Ge_2$, $YMn_{0.25}Ge_2$, $YFe_{0.3}Ge_2$, yielded the targeted products along with only very minor $YGe_{1.5+\delta}$ impurity. Transition metal-poor loadings produced impurity phase $Y_5Ge_3$ until reaching an optimal loading composition corresponding to the stable $YT_xGe_2$ phase. $YCo_{0.52}Ge_2$, and $YNi_{0.65}Ge_2$ yielded the most crystalline samples of the Co and Ni series; however, $Y_3Co_4Ge_{13}$ and $YNiGe_3$ were observed as by-products. $YGe_{1.5+\delta}$ was observed as a major impurity phase for transition metal-poor loadings, and remains present, but as a minor trace throughout the series, in the transition metal-rich samples. Table 1 summarizes the synthetic results in the phase analysis. Powder X-ray diffraction patterns of selected samples and $YGe_{1.5+\delta}$ are illustrated in Figure A1-5 and A7 in the appendix.

For the X-ray powder diffraction patterns, Le Bail refinements (Figure A6) were applied to fit all scale factors and lattice parameters, whereas the atomic sites and displacement parameters were later refined using the single crystal measurement data. The resulting profile residuals *Rp* varied between ca. 6.6-9.3% with weighted profile residuals $R_{wp}$ between ca. 9.0-13.0%. The refined lattice parameters for $BaCuSn_2$-$CeNi_{1-x}Si_2$-type $YT_xGe_2$ obtained from PXRD patterns show a ~4% increase along *b*-axis from Cr to Ni, which is consistent with the increasing occupancies on the T site. Additionally, the volume of the unit cells increase from Cr to Fe, Co, and Ni.

### 3.2 Structure Determination and Interpretation

Since there has been no reported data on the occupancy of the T site in the $YTGe_2$ (T= Fe, Co, and Ni) compounds from the last 30 years, with the exception of two phases $YCo_{0.55}Ge_2$ and $YMn_{0.3}Ge_2$, part of this work focused on the examination of the T site composition and occupancy.[14,15,21] The results confirmed the defects on the T site for all five $YT_xGe_2$ compounds. To gain further insight into crystal structures and chemical compositions of these $BaCuSn_2$-$CeNi_{1-x}Si_2$-type $YT_xGe_2$ phases, single crystal X-ray diffraction was carried out to extract accurate atomic positions. Single crystal XRD results are summarized in Tables 2 and 3. The anisotropic displacement parameters and bond distances are summarized in Tables S1 and S2, respectively. Figure 1 shows the agreement between experimental powder pattern and the calculated one obtained from the refined structure. All compounds crystallize in the orthorhombic structure with the space group *Cmcm* (No. 63) having four distinct 4*c* atomic sites. According to our refined structural data, two 4*c* sites are fully occupied by 8 Ge atoms, one 4*c* site is occupied by 4 Y atoms, and a forth 4*c* site accommodates transition metals. The T site is partially occupied by 22(1)% for Cr to 65(1)% for Ni. Elemental analysis is consistent with the refined occupancies with values of *x* being 0.18(3), 0.19(8), 0.27(3), 0.5(1) for Cr, Mn, Fe, and Co samples, respectively. Annealing of the samples at $900^{o}C$ for 5 days did not show evidence for any superstructure, in contrast with other $BaCuSn_2$-$CeNi_{1-x}Si_2$-type compounds.[23,24] Additionally, the SEM data (Figure A8) show the homogeneity of the sample.

Francois et al. gave a comprehensive structural description of the $RT_xGe_2$ (R= rare earth, T= 3*d* and 4*d* transition metals) as a comparison with the binary structure of $ZrSi_2$ and highlight the T atoms taking sites in the empty 4*c* square pyramidal sites created by the $RGe_8$ antiprism layers.[25,26] The insertion of T atoms does not result in any disturbance of the overall structure. To further understand the $YT_xGe_2$ structure, we initiated a tentative structural connection between superconductor $YGe_{1.5+\delta}$ and $BaCuSn_2$-$CeNi_{1-x}Si_2$-type $YT_xGe_2$ from the viewpoint of layered compounds, for example, transition metal dichalcogenides. While similar to the $ZrSi_2$ structure, the $YGe_{1.5+\delta}$ structure reported with space groups *Cmcm* or *Cmmm*, differs due to the $2_1$ glide symmetry transformation appearing along *b*-axis in *Cmcm*. [27,28] Traditionally, The Y and Ge atoms form Y@$Ge_5$ square pyramids along *b*-axis. As illustrated in Figure 2(*a*), alternatively, the ordered $YGe_{1.5+\delta}$ structure can be treated as a combination of three fragments, $Y_1Ge_1$ zigzag chains, the square planar $Ge_{0.5}$ (2Ge per layer /4Ge per unit cell) layer, and deficient $Ge_\delta$ ($0<\delta<1/2$) zigzag chain. Solely focusing on the structural fragments, $YT_xGe_2$ can be regarded in a similar way with three fragments, as shown in Figure 2(*b*). It consists of $Y_1Ge_1$ zigzag layers, the square planar $Ge_1$ layer, and deficient $T_x$ ($0<x<1$) zigzag layer (Figure 2(*c*)). Taking the chemical bonding interactions into consideration, the $BaCuSn_2$-$CeNi_{1-x}Si_2$-type $YT_xGe_2$ can be interpreted as $Y_1Ge_1$ plus $T_xGe_1$ zigzag layers.

### 3.3 Atomic Occupancies on the T Site

In an effort to figure out the factors that dictate the occupancies on the T site, we identified the relationship between atomic sizes of the transition metals and site occupancies. As mentioned previously, the unit cell dimensions are mostly dictated by Y and Ge as they create the original framework to which the transition metals are added. Figures 3(*a*) and 3(*b*) present the evolution of the site occupancy as a function of calculated atomic sizes and volume. One clear trend between theoretically atomic radius and occupancies can be observed in Figure 3(*a*); larger elements occupy a lower amount of the site, with almost the exponential dependence. Interestingly, the Figure 3(*b*) shows linear relationship between volume and occupancies for T= Cr, Fe, Co, and Ni except for Mn. Considering the magnetic properties, only $YMn_xGe_2$ exists the magnetic moments. It is highly possible that the magnetic spins on Mn atoms disturb the itinerant electron distribution and lead to the volume contraction in $YMn_xGe_2$. By the analysis of the atomic packing and transition metal-main group interactions in the structure of $YGe_{1.5+\delta}$ and $YT_xGe_2$ (Figure 2), a similar structural pattern is seen in other transition metal compounds.[29, 30] Considering that a large variety of compounds with very diverse compositions adopt the same structure, it seems clear that the structural stability and site occupancies of these compounds are heavily influenced by their valence electron concentrations or valence electron-to-atom ratios. The electron counts per metal M (M = Y+T) was determined using the equation [1×3 (e- per Y) + $x$×$T_m$ (e- per T) + 2×4 (e- per Ge)]/(1+$x$) with $T_m$ = 6 e- ~ 10 e-, leading to a range going from 10.1 to 10.6 e-/M. The electron count's range narrows significantly compared to hypothetical fully-occupied model of "$YTGe_2$" (8.5 e-/M for $YCrGe_2$ and 11.5 e-/M for $YNiGe_2$). Figure 3(*c*) shows the evolution of the occupancy as a function of the electron counts. One can observe two linear trends, the first one for Cr-Mn-Fe and the second one for Cr-Co-Ni. Compared to the reported data of other $RT_xGe_2$ (R = rare earth, T = Mn, Fe, Co and Ni) a narrow electron counts is seen for the compounds going from 9.8 e-/M for $LaFe_{0.69}Ge_2$ to 10.7 e-/M for $LuNi_{0.48}Ge_2$.[31] Additionally, the linear trend for the evolution of the occupancy as a function of the electron count was once again observed for the rare earth compounds (Figure 3(*b*) and A9). Thus, the atomic occupancies on T site are under the synergism of electron counts and atomic sizes. To deeply understand the chemical factors governing the structural stability, we propose a new concept-"valence electron density" to cooperate the electron counts and volumes. The basic idea is to generate the valence electron counts per unit cell, for example, the valence electron density of $YCr_xGe_2$ can be calculated using 10.1e- per $YCr_xGe_2$ × 4 $YCr_xGe_2$ per unit cell/261.13Å$^3$ = 0.155 e-/ Å$^3$. Similarly, we can obtain the valence electron density for T= Mn, Fe, Co, and Ni, which are all around 0.155 e-/ Å$^3$. In regards, the valence electron density could be another parameter to evaluate the chemical stability and transition metal occupancies.

### 3.4 Physical Properties

Figure 4(*a*) shows the temperature dependence of magnetic susceptibility ($\chi$) for $YCr_{0.22(1)}Ge_2$, $YFe_{0.282(6)}Ge_2$ and $YCo_{0.493(2)}Ge_2$ measured by applying 1 Tesla magnetic field. With decreasing temperature, $\chi$ for each compound increases without anomaly, indicating paramagnetic behavior. Fitting data using the modified Curie-Weiss law, $\chi = \chi_0 + \frac{C}{T-\theta}$ ($C = N_A\mu_0\mu_{eff}^2/3k_B$), we obtain Curie-Weiss temperature $\theta$, Curie constant C, and effective moment $\mu_{eff}$ for these compounds, as listed in Table 4. Both $\theta$ and C are small for $YCr_{0.22(1)}Ge_2$ and $YCo_{0.55(1)}Ge_2$, indicating little

magnetic interaction. For $YFe_{0.282(6)}Ge_2$, $\theta$ ~ -22 K, implying weak antiferromagnetic interaction. However, there is no long-range magnetic ordering. Intriguingly, the magnetic susceptibility of $YMn_{0.24(2)}Ge_2$ behaves differently from those compounds. Figure 4(*b*) shows the temperature dependence of $\chi$ for $YMn_{0.24(2)}Ge_2$ measured at 1 Tesla. There is a sharp raise near room temperature upon cooling, signature of magnetic ordering. The transition temperature $T_C$ ~ 293 K may be determined at the minimum of $d\chi/dT$, as shown in Fig. 4(*b*). Further isothermal hysteresis measurement at 2 K and 390 K, shown in Figure 4(*c*), indicates the ferromagnetic ordering below $T_C$. Using the Curie-Weiss formula to fit $\chi(T)$ above $T_C$, we obtain $\theta$ ~ 273 K, confirming the ferromagnetic interaction at high temperatures. From Curie constant C, the magnetic moment is estimated ~$3.6\mu_B$ per Mn, much higher than the ordered moment. This is likely due to the partial occupation of Mn leading to weak ferromagnetic interaction. Duraj et al. reported similar effective magnetic moment (~$3.5\mu_B$ per Mn), which orders antiferromagnetically at 395 K.[21] It is important to figure out whether discrepancy between our work and the previous report is solely due to small amount of Mn concentration. Obviously, introducing transition metals into the $YGe_2$ framework results in rather different magnetic properties from the Ge-deficient $YGe_{1.5+\delta}$, which is superconducting below 3.9 K as shown in Figure 4(*d*). It is likely that $YGe_{1.5+\delta}$ is a conventional superconductor, incompatible with magnetism. From the chemistry viewpoint, the additional $Ge_\delta$ atoms offer holes to the superconducting system. As transition metals are added to empty sites in the structure, this vacant $Ge_\delta$ layer is removed, leading more three dimensional structure. Such dimensionality transition and the reduced holes carrier concentration may be the reason for the suppression of superconductivity. The magnetic behavior observed for the $YT_xGe_2$ compounds highlights the key role the $Ge_\delta$ layer plays in the superconducting behavior of $YGe_{1.5+\delta}$.

## 4. Conclusions

$BaCuSn_2$-$CeNi_{1-x}Si_2$-type $YT_xGe_2$ (T= Cr, Mn, Fe, Co, and Ni) phases were synthesized and structurally characterized. Partial occupation is observed on the T site by the transition metal, and this is directly related to both atomic size and electron counts. The magnetic property measurements were collected for all $YT_xGe_2$ samples but the nickel product. Where T= Cr, Fe, and Co a paramagnetic behavior is observed whereas $YMn_{0.24(2)}Ge_2$ seems to be ferromagnetic with the transition temperature around 293 K. Replacing the deficient Ge layer with transition metals appears to suppress the superconductivity in $YGe_{1.5+\delta}$, leading to the discovery of ferromagnetic ordering in $YMn_{0.24(2)}Ge_2$ below 300 K.

## Acknowledgements

L.G., M.T.P., and W.X. deeply thank the support from Louisiana State University and the Louisiana Board of Regents Research Competitiveness Subprogram (RCS) under Contract Number LEQSF(2017-20)-RD-A-08 and the Shared Instrument Facility (SIF) at Louisiana State

University for the SEM-EDS. The work done by L.X. and R.J. was supported by the U.S. Department of Energy under grant No. DE-SC0016315.

## Appendix - Supplementary data

Tables of anisotropic temperature factors of $YT_xGe_2$ (T= Cr, Mn, Fe, Co, and Ni) single crystals, bond distances for $YT_xGe_2$ compounds, powder X-ray diffraction patterns of different loading composition for $YGe_{1.5+\delta}$ and $YT_xGe_2$ samples, Le Bail fit of $YT_xGe_2$ compounds, SEM images of $YT_xGe_2$ (T= Cr, Mn, Fe, and Co), evolution of the occupancy with respect to electron count per metal atom for $RT_xGe_2$ (R=Y, La-Lu; T=Mn, Fe, Co, and Ni).

**Table 1.** Loading compositions, phase analyses, unit cell parameters, and refined compositions for $YT_xGe_2$ samples (T = Cr, Mn, Fe, Co, and Ni) (PXRD = Powder X-ray Diffraction; SCXRD = Single-Crystal X-ray Diffraction)

| Loading composition (l.c.) x, from $YT_xGe_2$ | Phases (PXRD) | Unit cell parameters (Å) | | Composition (SCXRD) |
|---|---|---|---|---|
| | | PXRD | SCXRD | |
| Cr | | | | |
| 0.0625-0.125 | $YCr_xGe_2$, $YGe_{1.5+\delta}$ (majority phase) | - | - | - |
| **0.1875-0.25** | **$YCr_xGe_2$, $YGe_{1.5+\delta}$** | **a = 4.100(1)<br>b = 15.823(5)<br>c = 3.978(1)** | **a = 4.119(1)<br>b = 15.861(3)<br>c = 3.997(1)** | **$YCr_{0.22(1)}Ge_2$ (0.1875 l.c.)<br>$YCr_{0.23(1)}Ge_2$ (0.2 l.c.)** |
| 1 | $YCr_xGe_2$, $YGe_{1.5+\delta}$, $Cr_3Ge$ | - | - | - |
| Mn | | | | |
| **0.175-0.25** | **$YMn_xGe_2$, $YGe_{1.5+\delta}$** | **a = 4.1048(3)<br>b = 15.9100(9)<br>c = 3.9876(2)** | **a = 4.100(1)<br>b = 15.860(3)<br>c = 3.990(1)** | **$YMn_{0.24(1)}Ge_2$ (0.2 l.c.)<br>$YMn_{0.24(2)}Ge_2$ (0.25 l.c.)** |
| 0.3 | $YMn_xGe_2$, $Y_5Ge_3$, $YGe_{1.5+\delta}$ | - | - | - |
| Reference [[21]] | $YMn_xGe_3$ | - | - | $YMn_{0.3(1)}Ge_2$ (PXRD) |
| Fe | | | | |
| 0.175-0.2 | $YFe_xGe_2$, $Y_5Ge_3$, $YGe_{1.5+\delta}$ | - | a = 4.098(1)<br>b = 15.839(3)<br>c = 3.978(1) | $YFe_{0.28(1)}Ge_2$ (l.c. 0.175) |
| **0.28** | **$YFe_xGe_2$, $YGe_{1.5+\delta}$** | **a = 4.1151(2)<br>b = 15.8620(7)<br>c = 3.9923(1)** | **a = 4.097(1)<br>b = 15.839(3)<br>c = 3.977(1)** | **$YFe_{0.282(6)}Ge_2$** |
| Reference [[14]] | $YFe_xGe_2$ | a = 4.118<br>b = 15.903<br>c = 4.002 | - | - |
| Co | | | | |
| 0.15-0.2 | $YCo_xGe_2$, $Y_5Ge_3$, $YGe_{1.5+\delta}$ | - | - | - |
| 0.5 | $YCo_xGe_2$, $YGe_{1.5+\delta}$, unidentified phase | - | - | - |
| **0.52 - 0.55** | **$YCo_xGe_2$, $Y_3Co_4Ge_{13}$, $YGe_{1.5+\delta}$** | **a = 4.1150(2)<br>b = 16.0829(8)<br>c = 4.0193(1)** | **a = 4.103(1)<br>b = 16.034(3)<br>c = 4.006(1)** | **$YCo_{0.493(9)}Ge_2$ (0.52 l.c.)<br>$YCo_{0.55(1)}Ge_2$ (0.55 l.c.)** |
| Reference [[15]] | $YCo_{0.55}Ge_2$ | a = 4.106(6)<br>b = 16.12(1)<br>c = 4.026(4) | - | $YCo_{0.55(3)}Ge_2$ (PXRD) |
| Ni | | | | |
| 0.15-0.2 | $YNi_xGe_2$, $Y_5Ge_3$, $YGe_{1.5+\delta}$ | - | - | $YNi_{0.65(1)}Ge_2$ (0.2 l.c.) |
| **0.5-0.7** | **$YNi_xGe_2$, $YNiGe_3$, $YGe_{1.5+\delta}$** | **a = 4.1017(2)<br>b = 16.3320(7)<br>c = 4.0235(2)** | **a = 4.094(1)<br>b = 16.348(3)<br>c = 4.018(1)** | **$YNi_{0.658(7)}Ge_2$ (0.7 l.c.)** |
| Reference [[14]] | $YNi_xGe_2$ | a = 4.095(5)<br>b = 16.18(11)<br>c = 3.99(2) | - | - |

**Table 2.** Crystallographic data for $YT_xGe_2$ compounds (T = Cr, Mn, Fe, Co and Ni)

| Refined Formula | $YCr_{0.23(1)}Ge_2$ | $YMn_{0.24(2)}Ge_2$ | $YFe_{0.282(6)}Ge_2$ | $YCo_{0.493(9)}Ge_2$ | $YNi_{0.658(7)}Ge_2$ |
|---|---|---|---|---|---|
| F.W. (g/mol) | 245.92 | 247.82 | 249.87 | 263.41 | 272.69 |
| Space group; Z | *Cmcm* (No.63); 4 | | | | |
| *a* (Å) | 4.1190(8) | 4.1063(8) | 4.0974(8) | 4.1030(8) | 4.0940(8) |
| *b* (Å) | 15.861(3) | 15.863(3) | 15.839(3) | 16.034(3) | 16.348(3) |
| *c* (Å) | 3.9970(8) | 3.9571(8) | 3.9774(8) | 4.0060(8) | 4.0180(8) |
| V ($Å^3$) | 261.13(9) | 257.76(9) | 258.13(9) | 263.56(9) | 268.92(9) |
| Absorption Correction | Multi-scan | | | | |
| θ range (deg) | 2.568 – 33.079 | 2.568 – 33.210 | 2.572 – 33.112 | 2.541 – 33.162 | 2.492 – 33.124 |
| *hkl* ranges | $-4 \le h \le 6$<br>$-24 \le k \le 14$<br>$-5 \le l \le 6$ | $-6 \le h \le 6$<br>$-24 \le k \le 20$<br>$-4 \le l \le 6$ | $-6 \le h \le 5$<br>$-23 \le k \le 24$<br>$-6 \le l \le 6$ | $-4 \le h \le 6$<br>$-24 \le k \le 16$<br>$-6 \le l \le 5$ | $-5 \le h \le 6$<br>$-24 \le k \le 24$<br>$-6 \le l \le 6$ |
| No. reflections; $R_{int}$ | 1286; 0.1187 | 1291; 0.1068 | 1286; 0.0585 | 1374; 0.0893 | 1337; 0.0839 |
| No. independent reflections | 311 | 309 | 306 | 314 | 318 |
| No. parameters | 19 | 19 | 19 | 19 | 19 |
| $R_1$; $\omega R_2$ ($I>2\sigma(I)$) | 0.0489; 0.0974 | 0.0649; 0.0680 | 0.0304; 0.0568 | 0.0413; 0.0775 | 0.0378; 0.0680 |
| $R_1$; $\omega R_2$ (all $I$) | 0.1078; 0.1179 | 0.0835; 0.0928 | 0.0557; 0.0650 | 0.0941; 0.0958 | 0.0799; 0.0783 |
| Goodness of fit | 1.001 | 0.973 | 1.043 | 1.003 | 0.987 |
| Diffraction peak and hole ($e^-$/ $Å^3$) | 2.911; -2.422 | 2.366; -2.682 | 1.702; -1.361 | 3.485; -1.704 | 1.855; -1.918 |

**Table 3.** Atomic coordinates and equivalent isotropic displacement parameters of $YT_xGe_2$ compounds (T = Cr, Mn, Fe, Co, and Ni) ($U_{eq}$ is defined as one-third of the trace of the orthogonalized $U_{ij}$ tensor (Å$^2$)).

| Formula | Atom | Wyck. | *x* | *y* | *z* | $U_{eq}$ | Occ. |
|---|---|---|---|---|---|---|---|
| $YCr_{0.23(1)}Ge_2$ | Y | 4*c* | 0 | 0.1033(1) | ¼ | 0.0121(5) | 1 |
| | Cr | 4*c* | 0 | 0.3013(7) | ¼ | 0.005(3) | 0.23(1) |
| | Ge1 | 4*c* | 0 | 0.4483(1) | ¼ | 0.0140(6) | 1 |
| | Ge2 | 4*c* | 0 | 0.7479(1) | ¼ | 0.0322(7) | 1 |
| $YMn_{0.24(2)}Ge_2$ | Y | 4*c* | 0 | 0.8966(2) | ¼ | 0.0214(9) | 1 |
| | Mn | 4*c* | 0 | 0.3019(10) | ¼ | 0.015(5) | 0.24(2) |
| | Ge1 | 4*c* | 0 | 0.4483(2) | ¼ | 0.024(1) | 1 |
| | Ge2 | 4*c* | 0 | 0.7477(2) | ¼ | 0.039(1) | 1 |
| $YFe_{0.282(6)}Ge_2$ | Y | 4*c* | 0 | 0.1034(1) | ¼ | 0.0081(2) | 1 |
| | Fe | 4*c* | 0 | 0.3027(3) | ¼ | 0.008(1) | 0.282(6) |
| | Ge1 | 4*c* | 0 | 0.4481(1) | ¼ | 0.0105(3) | 1 |
| | Ge2 | 4*c* | 0 | 0.7478(1) | ¼ | 0.0260(4) | 1 |
| $YCo_{0.493(9)}Ge_2$ | Y | 4*c* | 0 | 0.1047(1) | ¼ | 0.0093(4) | 1 |
| | Co | 4*c* | 0 | 0.3119(3) | ¼ | 0.020(1) | 0.493(9) |
| | Ge1 | 4*c* | 0 | 0.4502(1) | ¼ | 0.0149(5) | 1 |
| | Ge2 | 4*c* | 0 | 0.7490(1) | ¼ | 0.0222(5) | 1 |
| $YNi_{0.658(7)}Ge_2$ | Y | 4*c* | 0 | 0.1057(1) | ¼ | 0.0078(3) | 1 |
| | Ni | 4*c* | 0 | 0.3173(2) | ¼ | 0.0124(8) | 0.658(7) |
| | Ge1 | 4*c* | 0 | 0.4534(1) | ¼ | 0.0129(4) | 1 |
| | Ge2 | 4*c* | 0 | 0.7492(1) | ¼ | 0.0177(4) | 1 |

**Table 4.** Magnetic data for $YT_xGe_2$ (T= Cr, Mn, Fe and Co) (PM=paramagnetic, FM=ferromagnetic, $\theta_C$ = Curie-Weiss temperature, C = Curie constant, $\mu_{eff}$ = effective magnetic moment)

| Compound | Behavior | $\theta_C$ (K) | *C*(emu•K/mol) | $\mu_{eff}$($\mu_B$ per formula) |
|---|---|---|---|---|
| $YCr_{0.23(1)}Ge_2$ | PM | -3.58 | $5.6\times10^{-4}$ | 0.07 |
| $YMn_{0.24(2)}Ge_2$ | FM | 273 | 0.093 | 0.86 |
| $YFe_{0.282(6)}Ge_2$ | PM | -22 | 0.003 | 0.15 |
| $YCo_{0.493(9)}Ge_2$ | PM | -0.25 | $8.1\times10^{-4}$ | 0.08 |

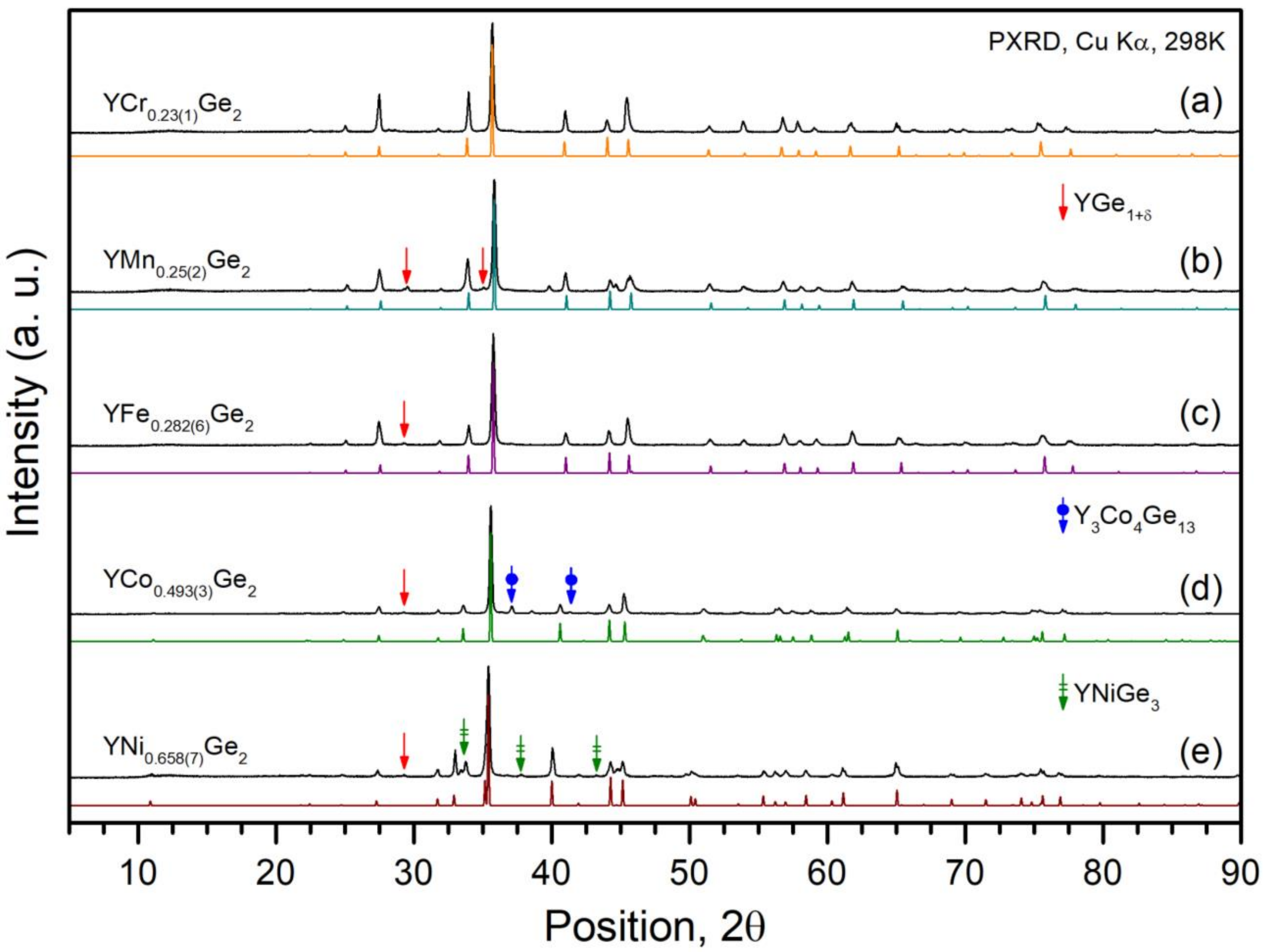

**Figure 1.** Comparison of experimental and calculated XRD patterns of $YT_xGe_2$ samples with T = (*a*) Cr, (*b*) Mn, (*c*) Fe, (*d*) Co and (*e*) Ni. Impurity peaks for $YGe_{1.5+\delta}$, $Y_3Co_4Ge_{13}$ and $YNiGe_3$ are labeled with red, blue and green arrows, respectively

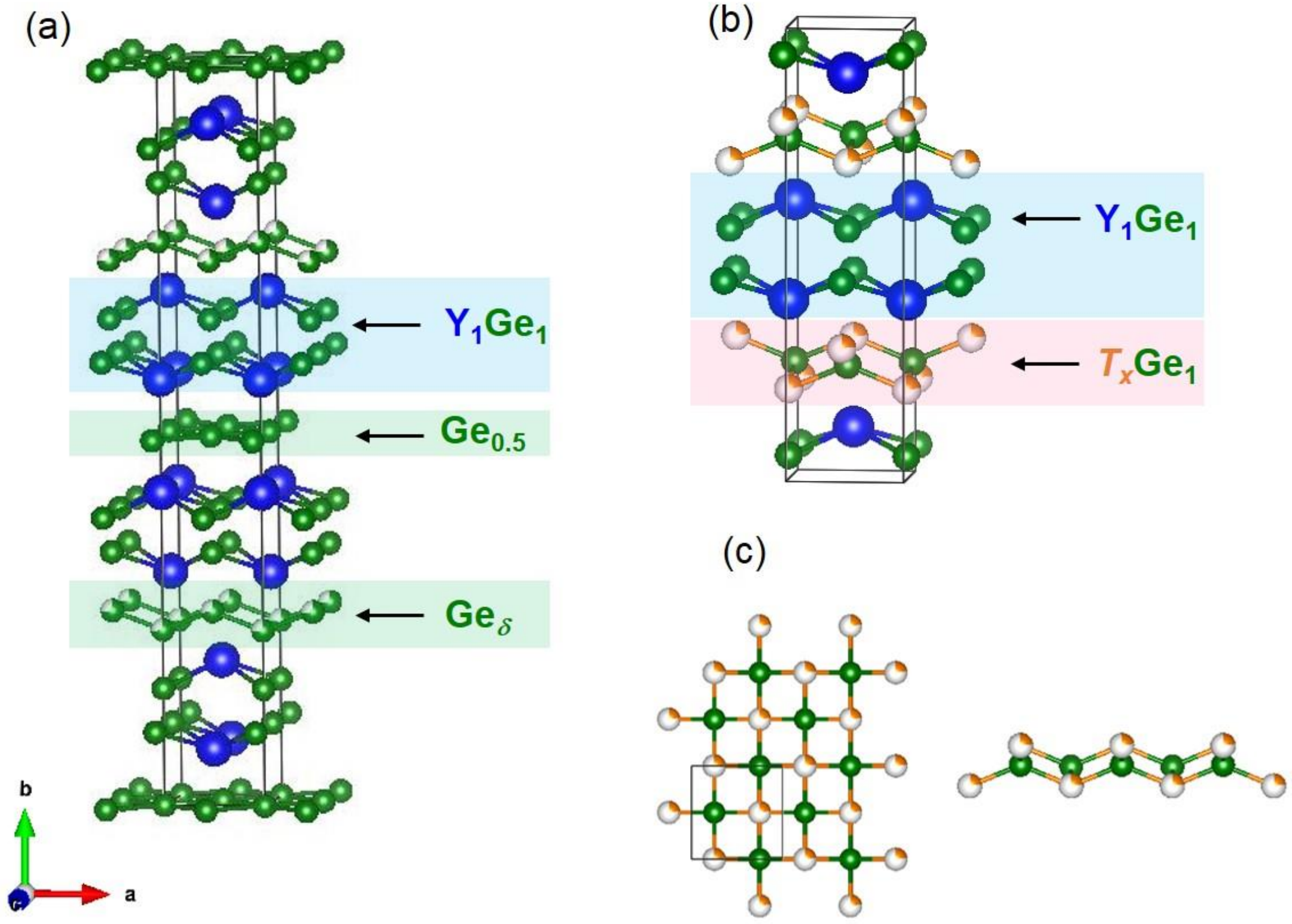

**Figure 2.** Crystal structures and atomic coordinate environments of (*a*) $YGe_{1.5+\delta}$ and (*b*) $YT_xGe_2$. Blue, green and orange spheres represent Y, Ge and T (Cr, Mn, Fe, Co, or Ni), respectively. (*c*) Top and side views of the $T_xGe_1$ layers in $YT_xGe_2$. The unit cell is traced in black.

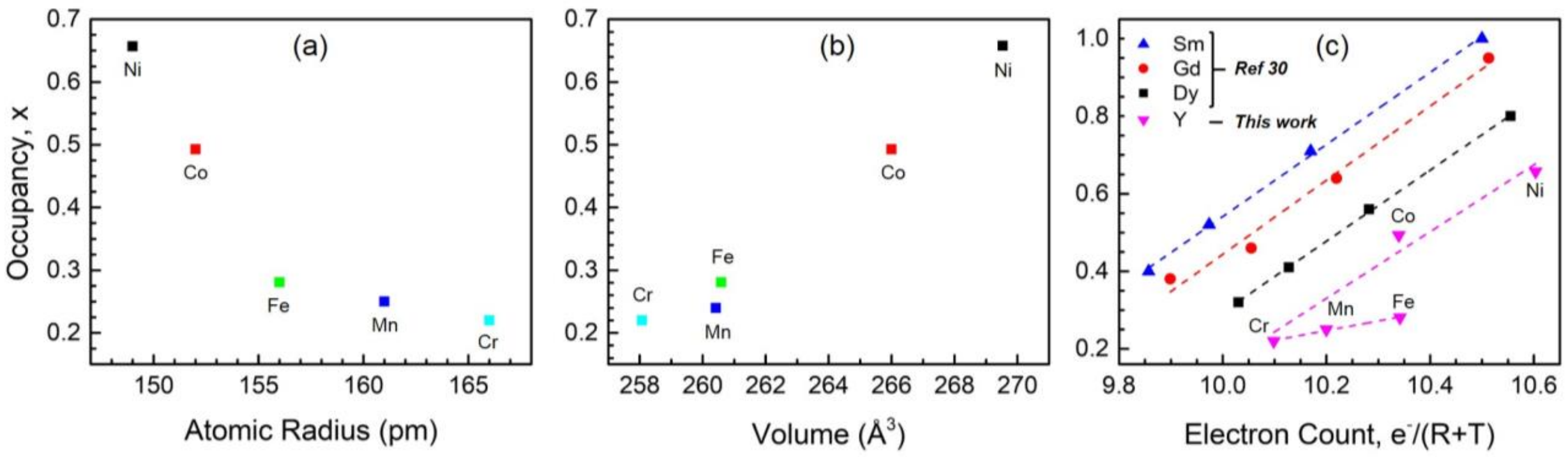

**Figure 3.** Evolution of the occupancy with respect to (*a*) atomic radius of the T transition metal in $YT_xGe_2$, (*b*) volume, and (*c*) electron count per metal atom (Y and T considered, in pink). The data for $RT_xGe_2$ where R= Dy, Gd, and Sm and T = Mn, Fe, Co and Ni (left to right) is shown for comparison; colored dash lines show the linear fitting for each series.

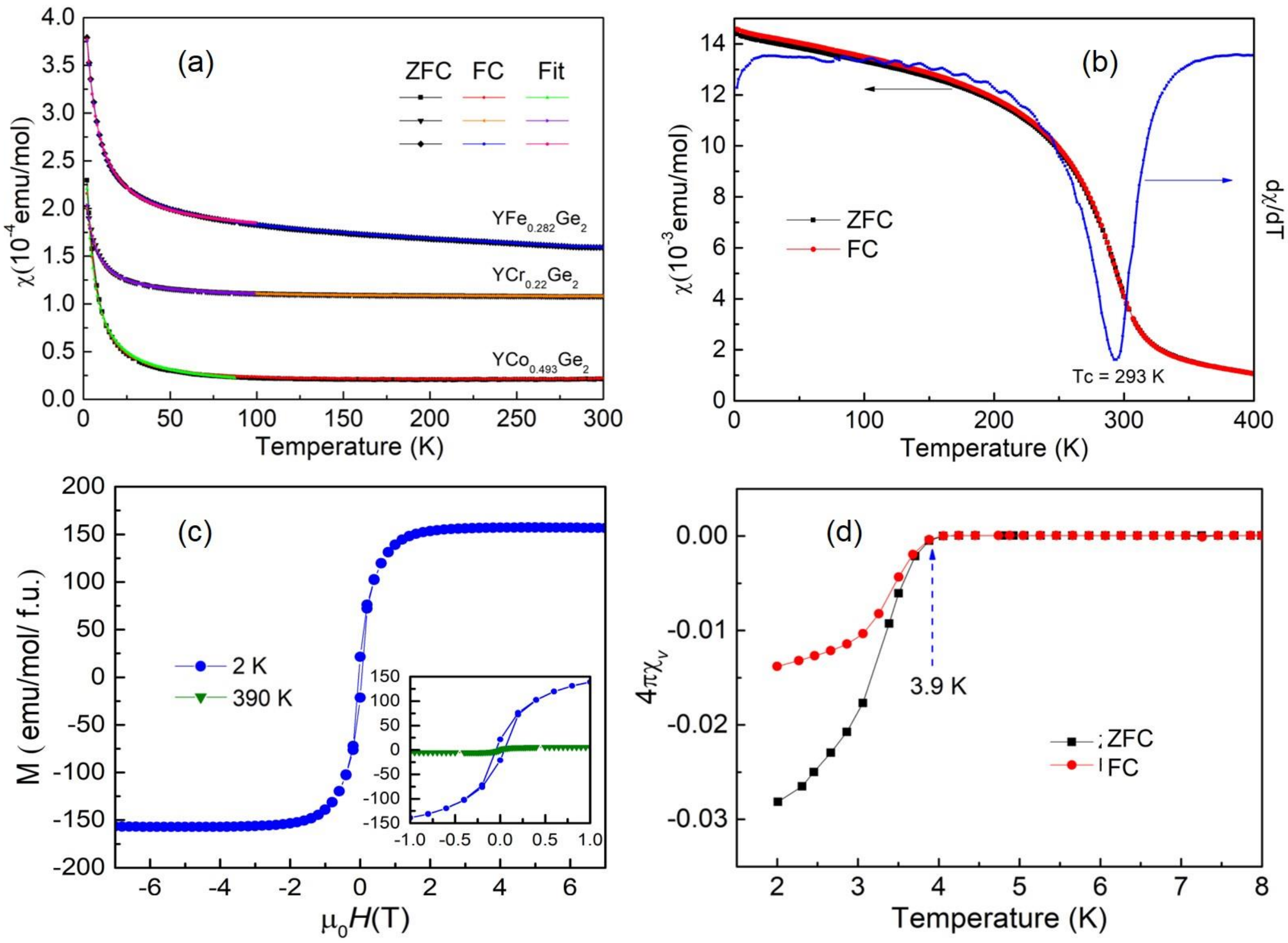

**Figure 4.** Magnetic measurements for $YT_xGe_2$ (T= Cr, Mn, Fe, and Co): (*a*) Temperature-dependent magnetization (ZFC-FC) under 1 T applied field for T = Cr, Fe, and Co. (*b*) Susceptibility and reciprocal susceptibility for T = Mn. The green line in shows the linear fit of the reciprocal susceptibility above room temperature. (*c*) Hysteresis curve at 2 K for $YMn_{0.24(2)}Ge_2$ showing ferromagnetic behavior. Inset shows the close up from -1 to 1 T in comparison with 390K. (*d*) Superconductive transition for $YGe_{1.5+\delta}$ in a 30 Oe field.

**For Table of Contents Only:**

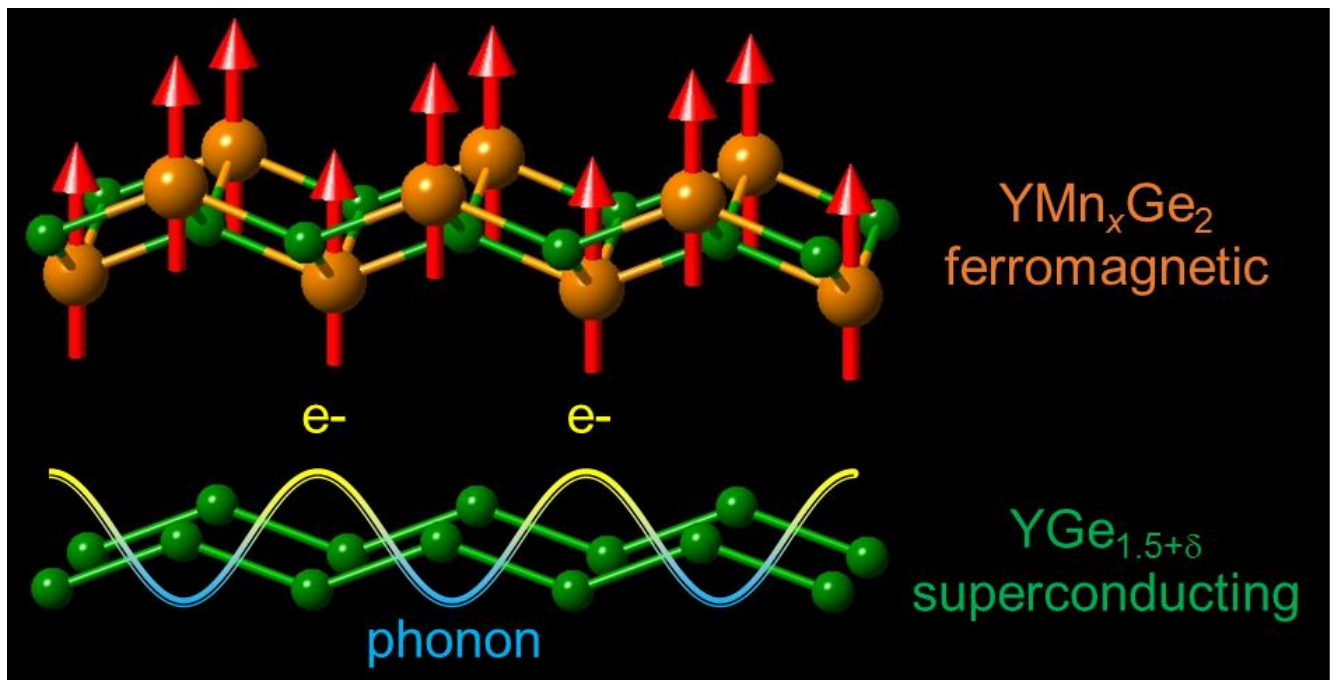

***Synopsis:*** Introducing 3*d* transition metals (T= Cr, Mn, Fe, Co, and Ni) into the Y-Ge system plays a critical role in suppressing superconductivity, providing great insights into the interplay between superconductivity and magnetism in layered intermetallics. In particular, the ferromagnetic moment arises for T = Mn with a Curie temperature of ~ 293 K and effective moment ~ 3.6 $\mu_B$/Mn.